\documentstyle[12pt,epsf]{article}
\begin{document}
{\bf On an Alternative Approach to Gravitation.  }
\begin{center} {\sl A.I.Nikishov}  \end{center}

{\sl P.N.Lebedev Institute of Physics, Russian Academy of Sciences, 117924
Moscow,Russia}
\begin{center}  Summary \end{center}

Severel energy-momentum "tensors" of gravitational field are considered and
compared in the lowest approximation. Each of them together with
energy-momentum tensor of point-like particles satisfies the conservation
laws when equation of motion of particles are the same as in general relativity.
It is shown that in Newtonian approximation the considered tensors differ
one from the other in the way their energy density is distributed between
energy density of interection (nonzero only at locations of particles)
and energy density of gravitational field.

Starting from Lorentz invariance the Lagrangians for spin-2, mass-0 field
are constracted. They differ only by divergences. From these Lagrangians by
Belinfante-Rosenfeld procedure the energy-momentum tensors are build. Only
one of them is suitable for explaining the perihelion shift. This tensor does
not coincide with Weinberg`s one (directly obtainable from Einstein equation).

It is noted that phenomenological field-theoretical approach (utilizing only
vertices and propagators) can lead to modification of theory in the region of
strong field, where till now no observational data are available.

 \section{ Introduction}

General relativity is a complete, elegant, and self-consistent theory.
Yet there is a necessity to obtain gravity by field-theoretical means starting
from flat spacetime, see e.g.[1-5].  It is widely believed that on this
way even dropping the general covariance requirement we naturally get
general relativity. It is supposed that in the lowest nonlinear
approximation this is demonstrated in detail by Thirring [2]. Yet this
conclusion can not be drawn from [2], see Sec.2.

In this connection and also because the gravitational collapse is
considered as the greatest crisis in physics [6] the research into possible
alternative theories acquire especial significance.
 It is quite natural to make the first step and to consider the simplest
 processes by utilizing vertices; the graviton propagator is known by
 analogy with electrodynamics.

 In the lowest nonlinear approximation it
 is necessary to know only 3-graviton vertex. We assume the simplest
 possibility: the source of graviton is the energy-momentum tensor of two other
 gravitons. In higher approximations probably other vertices will be needed.
Along this path one can find out what theories are possible without
assuming  general
covariance and a priori restriction on vertices. An important step in this
direction was made by Thirring [1-2].  We continue his investigation in the
 same approximation and restrict ourselves to point-like classical particels as
 sources of gravitation.  Mainly we are interested in the  simplest system
consisting of a Newtonian center and test particle moving in its field.

 In general relativity classical particles move along geodesics in Riemannian
 space. This is the incarnation of  equivalence principle and it is more
 reliable than specific equation determining the gravitational field [9].
Besides, the equation of motion for particles are obtainable from variational
 principle without any appeal to general relativity, see equation (2) on page
 181 in [6].  It is easy to verify also that eq.(11) in [2], derived in this
way coincide exactly with that of general relativity. As to the equation
 determining gravitational field, it is possible to think that the
 phenomenological field-theoretical approach will lead to more complicated
 algorithm for getting the field. An interesting possibility in this direction
 was pointed out  by Schwinger [10], see also Sec. 6.

 It is reasonable to believe together with Einstein that for some reason or
 other the singular behaviour near the gravitational radius does not
 correspond to reality, see \S 15 in  €.Pais`s book [11] and Einstein paper
[12].  .  At present the Schwarzschild singularity is concidered as fictitious
 by many researchers because the geometry is nonsingular there. See however the
 text after (2.2.6) in [13] and after eq.(9.40) in [14], where they say
convincingly
 about  {\sl physical} singularity. By field-theoretical approach it is
 difficult to understand why in a finite system the acceleration of a
 freely falling particle becomes unlimited when it nears the horizon. Such a
behaviour should be connected with the fact that according to [15] the
gravitational energy in vacuum outside the sphere of radius R goes to
  $-\infty $ for $ R\to r_g $. In conformity with this the energy of matter
 and gravitational field inside the sphere of radius $R$ goes to $+\infty $,
in such away that total energy of spherical body is equal to its mass.
But if a theory predict that the absolute value of field energy outside sphere
of radius R might be greater than total energy of a body then
the analogy with electrodynamics suggests that the concept of external field
becomes inapplicable [16].
The belief in general relativity in similar circumstances is based upon the
concept of nonlocalazability of gravitational energy, see, e.g. \S 20.4 in [6].
What is more, general relativity does not need as a rule the gravitational
energy -momentum pseudotensor.

 The situation changes drastically, when we begin to constract gravity theory
 starting from flat spacetime. Here the gravitational energy-momentum tensor
  appears to be necessary to describe the 3-graviton interaction.
The nonlinear correction to the motion of a test particle depends on the chosen
 energy-momentum tensor.  The latter is build from field Lagrangian, which is
 not unique as one can add to it some divergence terms. This leads to different
energy-momentum tensors. They can give rise to gravitational energy densities,
which may have even different signs. The question of sign of energy density
is of interest by itself. Provided the sign turns out to be positive, one
should expect the weakening of gravitational interaction at $r\sim r_g=2GM $
in comparison with Newtonian one in order that the gravitational energy
outside the sphere of radius $r$ were much less than the mass $M$ of the
center. The possibility of decreasing of interaction at small distances is
suggested also  by the behaviour of attraction force between two bodies
supported by Weil`s strut, see,  \S 35 ¢ [17].

In order to understand in what way the various energy-momentum tensors
differ one from the other we consider the following tensors: Thirring`s
 [1,2], Landau-Lifshitz`s [16], Papapetrou-Weinberg`s [9] and tensor obtained
from the Lagrangian given in Exercise 7.3 in [6].  The second and third tensors
are representatives of general relativity, the rest are build from Lagrangians
of free field of spin-2, mass-0 particles and symmetrized by the
Belinfante-Rosenfeld procedure, see, \S 1 Ch.7 in [18].

In the considered approach it is suitable to subdivide the 3-graviton vertex
into three vertices in accordance with three possibilities for choosing
two gravitons out of three to form the  energy-momentum tensor,- the source
for the third graviton, see Sec.3. So three diagrams contribute to nonlinear
correction to the field. In contrast with this the energy-momentum tensor,
figuring in the solution of Einstein equation by iteration procedure, is so
defined that the correction to field is given simply by means of propagator,
i.e. by one diagram only, see Sec.2.

The main result of the paper is this: starting from quadratic Lagrangians
(differing by divergence terms) of spin-2, mass-0 particles, the
energy-momentum tensors are constructed by Belinfante-Rosenfeld procedure.
It turns out that only certain combination of these tensors is fitted for
correct description of perihelion shift. This combination does not coincide
with Papapetrou-Weinberg tensor.

The investigation of possibilities of phenomenological approach to gravitation
without use of general covariance  seems to us very promising. Valuable
undertaking in this direction was made in [19].

\section { Thirring`s energy-momentum tensor }
 Throughout the paper we use
$$
g_{\mu\nu}=\eta_{\mu\nu}+h_{\mu\nu}   \eqno (1)
$$
 In Sections 2, 3, 5 $ \eta_{\mu\nu}=diag(1,-1,-1,-1,)$; in Sections 4 and 6
 $ \eta_{\mu\nu}=diag(-1,1,1,1,)$. In this Sec. we use Thirring`s notation
 [2]; both greek and latin indices run from 0 to 3.

The gravitational field is described by the symmetric tensor $h_{\mu\nu}$,
which contains spin-2 and lower spins, see, e.g. [3]. The unnecessary spins
 (spin-1 and one of spin-0) are excluded by Hilbert gauge :
 $$
{\bar h^{\mu\nu}}{}_{,\nu}\equiv
(h^{\mu\nu}-\frac12\eta^{\mu\nu}h)_{,\nu}=0,\quad h={h^{\sigma}}_{\sigma},\quad
h_{,\nu}\equiv\frac{\partial}{\partial x^{\nu}}h.  \eqno (2)
   $$
   One way to
obtain Thirring`s tensor is to start from general relativity Lagrangian
$\sqrt{-g}R$. If we remove terms with second derivatives of $g_{\mu\nu}$ into
divergence terms and drop the latter, we get the function $ G(x) $ in eq.(93.1)
 in [16]. Retaining in it only quadratic in $h_{\mu\nu}$ terms we get $$
  G(x)=\frac14[h_{\mu\nu,\lambda}h^{\mu\nu,\lambda}-
2h_{\mu\nu,\lambda}h^{\lambda\nu,\mu}+2{h_{\nu\mu}}^{,\mu}h^{,\nu}-
h_{,\lambda}h^{,\lambda}] \eqno (3)
$$
This is equivalent to Thirring`s Lagrangian [2]
$$
\stackrel{f}{L}=\frac12[\psi_{\mu\nu,\lambda} \psi^{\mu\nu,\lambda}-
2\psi_{\mu\nu,\lambda}\psi^{\lambda\nu,\mu}+2{\psi_{\mu\nu}}^{,\mu}\psi^{,\nu}-
\psi_{,\lambda}\psi^{,\lambda}]. \eqno (4)
$$
Here
$$
\psi_{\mu\nu}=-h_{\mu\nu}/2f, \quad \psi={\psi_{\sigma}}^{\sigma},
 \quad f^2=8\pi G,\quad  G=6.67\cdot10^{-8}cm^{3}/g\cdot sec^{2.}  \eqno (5)
$$

Using  $\psi_{\mu\nu}$ instead of $h_{\mu\nu}$ is justified because then the
analogy with electrodynamics becomes more close:  $\psi_{\mu\nu}$ is analogous
to vector- potential $A_{\mu}$ and has the same dimensionality, $M\sqrt G$
has the dimensionality of electromagnetic charge. We note that the Lagrangian
(4) exactly corresponds to Schwinger`s Lagrangian [10], who uses the notation
$\eta_{\mu\nu}=diag(-1,1,1,1,)$ and $2h^{Sch}_{\mu\nu}=-h_{\mu\nu}^T=
-h_{\mu\nu}.$

The canonical energy-momentum tensor following from (4), has the form
\setcounter{equation}{5}
\begin{eqnarray}
\stackrel{f}{T^{\gamma\delta}}
=\varphi^{\mu\nu,\delta}{\varphi_{\mu\nu}}^{,\gamma}
-\frac12\varphi^{,\delta}\varphi^{,\gamma}-
2\varphi^{\mu\nu,\delta}{\varphi^{\gamma}}_{\nu,\mu}-
\eta^{\gamma\delta}\stackrel{f}{L},\nonumber\\
\stackrel{f}{L}=\frac12[\varphi_{\mu\nu,\lambda}\varphi^{\mu\nu,\lambda}-
\frac12\varphi_{,\lambda}\varphi^{,\lambda}-
2\varphi_{\mu\nu,\lambda}\varphi^{\lambda\nu,\mu}], 
\end{eqnarray}

$$
\varphi_{\mu\nu}\equiv\bar\psi_{\mu\nu}=\psi_{\mu\nu}-\frac12\eta_{\mu\nu}\psi,
\quad\psi={\psi_{\sigma}}^{\sigma}.\eqno (7)
$$
Using $\varphi_{\mu\nu}$ instead of $\psi_{\mu\nu}$ is handy as many
expressions become more compact and the consequences of imposition of Hilbert
gauge more clear.

The energy-momentum tensor for a static point-like mass  (Newtonian center)
 $$
\stackrel{M}{T_{\mu\nu}}=M\delta(\vec x)\delta_{\mu 0}\delta_{\nu 0}.\eqno (8)
$$
In linear approximation this source generate the field
$$
  \varphi_{\mu\nu}=-\bar h_{\mu\nu}/2f=
\frac{fM}{4\pi\vert\vec x\vert}\delta_{\mu 0}\delta_{\nu 0},\quad
\bar h_{\mu\nu}=4\phi\delta_{\mu o}\delta_{\nu 0}, \eqno (9)
$$
satisfying Hilbert condition (2). For one Newtonian center $\phi=-GM/r$. For
several centers
$$
\phi=-G\sum_a\frac{m_a}{\vert\vec r-\vec r_a\vert}. \eqno (10)
$$
In terms of
$$
h_{\mu\nu}=\bar h_{\mu\nu}-\frac12\eta_{\mu\nu}\bar h,
\quad\bar h=\bar{ h_{\sigma}}^{\sigma},
\eqno (11)
$$
we have
$$
h_{\mu\nu}=2\phi\delta_{\mu\nu},\quad h={h_{\sigma}}^{\sigma}=-4\phi=-\bar h.
\eqno (12)
$$
The energy density of field (9) is positive
$$
\stackrel{f}{T}{}^{00}=\frac{1}{8\pi G}(\nabla\phi)^2.\eqno (13)
$$
But $\stackrel{f}{T^{\gamma\delta}}$ ought to be supplemented to symmetric
one by the spin part:
$$
\theta^{\gamma\delta}=\stackrel{f}{T}{}^{\gamma\delta}+
\stackrel{s}{T}{}^{\gamma\delta}.
\eqno (14)
$$
For Newtonian center Thirring obtains
$$
\stackrel{s}{T^{\gamma\delta}} =
-\frac{1}{\pi G}(\nabla\phi)^2\delta_{\gamma 0}\delta_{\delta 0}  \eqno (15)
$$
So in this case $\theta^{00}$ is negative
 $$
 \theta^{00}=-\frac{7}{8\pi G}(\nabla\phi)^2.\eqno (16)
 $$

 Turning now to conservation laws of total energy-momentum we remind first
how matters stand in general relativity. There the energy-momentum tensor
of point-like particles $\stackrel{p}{T^{\mu\nu}}$ is connected with its
counterpart in special relativity $\cal T^{\mu\nu}$ by the relation, see
 (33.4), (33.5) and (106.4) in [16]:
  $$
\sqrt{-g}\stackrel{p}{T^{\mu\nu}}={\cal T}^{\mu\nu}=
\sum_am_au^{\mu}u^{\nu}\frac{ds}{dt}\delta(\vec x-\vec x_a(t)),\quad
u^{\mu}=dx^{\mu}/ds, \eqno (17)
$$
$g$ is determinant of $g_{\mu\nu}$. In terms of  $\cal T^{\mu\nu}$ the
conservation laws are (see
(96.1) in [16])
 $$
 {{\cal T}^{\mu}}_{\nu,\mu}=[{\cal
T}^{\mu\tau}(\eta_{\tau\nu}+h_{\tau\nu})]_{,\mu}= \frac12h_{\mu\sigma,\nu}{\cal
T}^{\mu\sigma}. \eqno (18)
 $$
We shall see below that ${\cal T}^{\mu\tau}h_{\tau\nu}$ can be interpreted as
(part of)  interaction energy-momentum tensor.

As is known the equation of motion of particles in general relativity
$$
\frac{d^2x^k}{ds^2}+\Gamma^k_{mj}u^mu^j=0 \eqno (19)
$$
 is contained in consevation laws.
Indeed from
 $$
\stackrel{p}{T}{}^{jk}{}_{;j}=\stackrel{p}{T}{}^{jk}{}_{,j}+
\Gamma^j_{mj}\stackrel{p}{T}{}^{mk}+\Gamma^k_{mj}\stackrel{p}{T}{}^{jm}=0
 \eqno (20)
 $$
 taking into account that from definition of $\stackrel{p}{T}{}^{jk}$ in (17)
$$
\stackrel{p}{T}{}^{jk}{}_{,j}=-\frac12(-g)^{-\frac32}(-g)_{,j}{\cal T}^{jk}+
(-g)^{-\frac12}{\cal T}^{jk}{}_{,j},\quad \Gamma^j_{mj}=\frac{1}{2g}g_{,m}{},
\eqno (21)
$$
we get
$$
{{\cal T}^{jk}}_{,j}+\Gamma^k_{mj}{\cal T}^{mj}=0. \eqno (22)
$$
This is equivalent to (19), because [9]
$$
{{\cal T}^{jm}}_{,j}=\sum_a\frac{dp^m}{dt}\delta(\vec x-\vec x_a(t)).\eqno (23)
 $$
Going back to field-theoretical approach, we rewrite the equation of motion of
particles (19) in the lowest approximation
 $$
\frac{du^{\mu}}{ds}=\frac{d^2x^{\mu}}{ds^2}=
\frac{1}{2}h_{\alpha\beta}{}^{,\mu}u^{\alpha}u^{\beta}-
h^{\mu}{}_{\alpha,\beta}u^{\alpha}u^{\beta}. \eqno (24)
$$
Just at such movement of particles the divergence of total energy-momentum
ought to be zero, and inversely, from zero divergence follows eq. (24). From
(23) and (24) we find
 $$ {\cal T}^{\gamma\delta}{}_{,\gamma}=
\frac12h_{\alpha\beta}{}^{,\delta}{\cal T}^{\alpha\beta}-
h^{\delta}{}_{\alpha,\gamma}{\cal T}^{\alpha\gamma}\quad, \eqno (25)
 $$
  This agrees with (22) in considered appoximation. With the same accuracy
this can be rewritten as
  $$
 ({\cal T}^{\gamma\delta}+ {\cal T}^{\gamma\alpha}
h_{\alpha}{}^{\delta})_{,\gamma}=
\frac12h_{\alpha\beta}{}^{,\delta}{\cal T}^{\alpha\beta}. \eqno (26)
$$

   Using linearized Einstein equation for $\varphi_{\mu\nu}$
 \setcounter{equation}{26}
 \begin{eqnarray}
\varphi_{\mu\nu,\lambda}{}^{\lambda}-
\frac12\eta_{\mu\nu}\varphi_{,\lambda}^{}\lambda-
\varphi^{\lambda}{}_{\nu,\mu\lambda}-\varphi^{\lambda}{}_{\mu,\nu\lambda}=
f\bar{\cal T}{}_{\mu\nu}\quad,   \nonumber\\
\bar{\cal T}{}_{\mu\nu}={\cal T}_{\mu\nu}-\frac12\eta_{\mu\nu}{\cal T}\quad,
\quad{\cal T}={\cal T}_{\sigma}{}^{\sigma} \quad,
\end{eqnarray}
we get
$$
\stackrel{f}{T}{}^{\gamma\delta}{}_{,\gamma}=
f\varphi^{\alpha\beta,\delta}\bar{\cal T}_{\alpha\beta}=
-\frac12\bar h_{\alpha\beta}{}^{,\delta}\bar{\cal T}^{\alpha\beta}=
-\frac12h_{\alpha\beta}{}^{,\delta}{\cal T}^{\alpha\beta}. \eqno (28)
$$
 Spin part of energy-momentum tensor is conserved by itself and do not
contribute to conservation laws :
$$
\stackrel{s}{T}{}^{\gamma\delta}{}_{,\gamma}=0.\eqno (29)
$$

 So from (26) and (28) it follows that  the conserved tensor contains
in itself the interaction tensor [2]
 $$ \stackrel{int}{T}{}^{\gamma\delta}= {\cal
T}^{\gamma\alpha}h_{\alpha}{}^{\delta}. \eqno (30)
 $$

  But it is not symmetric. In order to understand the reason we have to
consider the properties of $\stackrel{s}{T}{}^{\gamma\delta}$ in detail
despite the fact that it does not take part in conservation laws written in the
form of eqs.(26) and (28). According to known rules   [2,18] we have
 $$
\stackrel{s}{T}{}^{jk}=-F^{jik}{}_{,i}-F^{kij}{}_{,i}-F^{ikj}{}_{,i}\quad,
\eqno (31)
$$
$$
F^{jik}=\frac{\partial L}{\partial\varphi_{\alpha\beta,j}}%
(\varphi^k{}_{\alpha}\eta^i{}_{\beta}-\varphi^i{}_{\alpha}\eta^k{}_{\beta}),
 \qquad L = \stackrel{f}{L}.
\eqno (32)
$$
 The antisymmetric part of (31) is contained only in the last term. For it we
have
 $$
 -F^{ikj}{}_{,i}=\left(\frac{\partial
L}{\partial\varphi_{\alpha j,i}}\right)%
_{,i}\varphi^k{}_{\alpha}-\left(\frac{\partial L}{\partial\varphi%
 _{\alpha%
k,i}}\right)_{,i}\varphi^j{}_{\alpha}+\frac{\partial L}%
{\partial\varphi_{\alpha j,i}}\varphi^k{}_{\alpha,i}-
\frac{\partial L}{\partial\varphi_{\alpha k,i}}\varphi^j{}_{\alpha,i}.\eqno (33)
$$
  The first two terms on the right hand side symmetrize the interaction tensor,
the last two terms symmetrize the canonical one.

It is not seen directly from  (6) and (31) that field energy-momentum tensor
 $\theta^{\gamma\delta}$ in (14) is symmetric. This agrees with the fact that
 the proof of symmetry utilizes the Euler-Lagrange equations for field which
 is considered as free [18]. We are interested in interacting field. So using
 linearized Einstein equation (27) with source, we get
\setcounter{equation}{33}
 \begin{eqnarray} \left(\frac{\partial
L}{\partial\varphi_{\alpha j,i}}\right)_{,i}%
 \varphi^k{}_{\alpha}-\left(\frac{\partial L}{\partial\varphi_{\alpha k,i}}%
\right)_{,i}\varphi^j{}_{\alpha}=f(\bar{\cal T}{}^{\alpha
j}\varphi^k{}_{\alpha} -\bar{\cal T}{}^{\alpha k}\varphi^j{}_{\alpha})
\nonumber \\ =f({\cal T}{}^{\alpha j}\psi^k{}_{\alpha}-{\cal T}{}^{\alpha k}%
\psi^j{}_{\alpha})=\frac12({\cal T}{}^{\alpha k}h^j{}_{\alpha}- {\cal
T}{}^{\alpha j}h^k{}_{\alpha}).
  \end{eqnarray}
In the last equation we use the connection between $\psi_{\mu\nu}$ and
$h_{\mu\nu}$, see (5).  Substituting in (30) $\gamma \to j, \delta \to k$,  we
see that the sum of (30) and (34) is symmetric. This result retains if we start
 from another Lagrangian differing from Thirring`s one in (6) by divergence
because the linearized equation remains the same.

One should take into account however that the symmetric part of
$\stackrel{s}{T}{}^{jk}$ can also contain terms of interaction type. So for
the Lagrangian in (6) similarly to (34) we find
 \setcounter{equation}{34}
\begin{eqnarray}
 -F^{jik}{}_{,i}-F^{kij}{}_{,i}=[f{\cal
T}-2\varphi_{il}{}^{,il}]\varphi^{kj}+ f[\bar{\cal T}{}^{\alpha
j}\varphi^k{}_{\alpha}+\bar{\cal T}{}^{\alpha k}%
\varphi^j{}_{\alpha}]\nonumber\\+2(\varphi^{ij,\alpha}{}_i\varphi^k{}_{\alpha}
+\varphi^{ki,\alpha}{}_i\varphi^j{}_{\alpha})+
2\varphi^{j\alpha,i}\varphi^k{}_{\alpha,i}-(\varphi^{\alpha i,j}%
\varphi^k{}_{\alpha,i}+\varphi^{\alpha i,k}\varphi^j{}_{\alpha,i})\nonumber\\
-2(\varphi^{jk,\alpha}{}_i\varphi^i{}_{\alpha}+\varphi^{jk,\alpha}%
\varphi^i{}_{\alpha,i})+2\varphi^{ki,\alpha}\varphi^j{}_{\alpha,i}.
\end{eqnarray}
As a result we get for $\stackrel{s}{T}{}^{jk}$
\setcounter{equation}{35}
\begin{eqnarray}
 \stackrel{s}{T}{}^{jk}=2[(\varphi^{ij,\alpha}{}_i\varphi^k{}_{\alpha}+
\varphi^{ik,\alpha}{}_i\varphi^j{}_{\alpha})-\varphi^{i\alpha}{}_{,i\alpha}%
\varphi^{kj}
+\varphi^{j\alpha,i}\varphi^k{}_{\alpha,i}-\varphi^{jk,\alpha}{}_i%
\varphi^i{}_{\alpha}\nonumber\\
-\varphi^{jk,\alpha}\varphi^i{}_{\alpha,i}
+\varphi^{ki,\alpha}\varphi^j{}_{\alpha,i}]-2\varphi^{i\alpha,j}%
\varphi^k{}_{\alpha,i}+2f{\cal T}{}^{j\alpha}\varphi^k{}_{\alpha}.
\end{eqnarray}
Here last but one term, added to $\stackrel{f}{T}{}^{jk}$, makes it symmetric.
The last term can be rewritten in terms of $h_{\mu\nu}$ in the form, see (9)
and (11), (30),
 $$
 -{\cal
T}{}^{j\alpha}(h_{\alpha}{}^k-\frac12\eta_{\alpha}{}^kh)=
-\stackrel{int}{T}{}^{jk}+\frac12{\cal T}{}^{jk}h. \eqno (37)
 $$
So the symmetrization of $\stackrel{int}{T}{}^{jk}$ in (30) is reduced to its
replacement by $\frac12{\cal T}{}^{jk}h$. This tensor is nonzero only where
particles are present. For Newtonian centers the corresponding energy density
 $$
 \frac12{\cal T}{}^{00}h=-2{\cal
T}{}^{00}\phi \eqno (38)
 $$
is positive (contrary to our intuition and) contrary to
$\stackrel{int}{T}{}^{00}$ in (30), see (12) and (10), where $h$ and $\phi$
are given for Newtonian centers.

We note that the use of linearized Einstein eq. (27) in the expression for
$\stackrel{s}{T}{}^{jk}$ leads to that eq. (29) is satisfied only with
considered accuracy. The presence of interaction energy-momentum tensor means
the appearence of such vertex:  the energy-momentum tensor of matter
together with one of gravitons serves as a source for other graviton, see Fig.1

Thirring assumes that his tensor $\theta^{\gamma\delta}$ (see (14), (6), (31))
 is an analog of energy-momentum tensor
figuring in the r.h.s. of Einstein equation when iteration procedure is used.
In other words the nonlinear correction to field is given only by diagram (2a)
in Fig.2. On this Fig. the short straight line has only conditional meaning:
it represents the source of gravitons, namely the energy-momentum tensor build
of two gravitons (real or virtual) shown on Fig.2 as joining the ends of this
line. The graviton emerging from the middle of straight line is emitted or
absorbed by this source. On diagram (2a) the energy-momentum tensor is build
from gravitons of Newtonian center. On diagrams (2b) and (2c) one of the
virtual gravitons of Newtonian center interact with energy-momentum tensor of
two other gravitons. All three diagrams on Fig.2 correspond to one Feynman
diagram obtained by contracting the short straight line to a point.

The contribution to nonlinear correction for field from diagram (2a) is easy to
obtain. Indeed, from  (14), (6) and (15) we have
 $$
\theta^{jk}=\stackrel{f}{T}{}^{jk}=\frac{GM^2}{4\pi}\left(\frac{x^jx^k}{r^6}
-\frac12\frac{\delta_{jk}}{r^4}\right),\quad j,k=1,2,3.\eqno (39)
$$
Using now the field equation in Hilbert gauge with  $\theta^{\mu\nu}$ from (16)
and (39)
 $$
 \Box\bar h^{\mu\nu}=-16\pi
G\theta^{\mu\nu},\quad\Box=\frac{\partial^2}{\partial t^2}-\nabla^2 \eqno
(40)
 $$
we find
 $$
 \bar h^{00}=-7\phi^2,\quad\bar h^{ik}=-\frac{G^2M^2}{r^4}x^ix^k,\quad\phi=
-\frac{GM}{r},\quad i,k=1,2,3. \eqno (41)
$$
   Here easily verifiable relations
$$
\nabla^2\frac{x^ix^k}{r^4}=\frac{2\delta_{ik}}{r^4}-\frac{4x^ix^k}{r^6},
\quad \nabla^2\frac{1}{r^2}=\frac{2}{r^4} \eqno (42)
$$
 were used. The obtained $\bar h^{\mu\nu}$ satisfies Hilbert condition.
Going over to $h_{\mu\nu}=\bar
h^{\mu\nu}-\frac12\eta_%
 {\mu\nu}\bar h$, we find the following nonlinear
corrections
 $$
 h_{00}=-4\phi^2,\quad
h_{ik}=-G^2M^2(\frac{x_ix_k}{r^4}+\frac{3\delta_{ik}}%
 {r^2}). \eqno (43)
 $$
Index 2 in  $\stackrel{(2)}{h}{}_{\mu\nu}$, indicating the order of correction
in powers of  $G$, is dropped for brevity.

Finally from (12) and (43) we have
  $$
ds^2=g_{00}dt^2-(1-2\phi+3\phi^2)\delta_{ik}dx^idx^k-\phi^2\frac{x_ix_k%
dx^idx^k}{r^2}, \eqno (44)
$$
$$
g_{00}=1+2\phi-4\phi^2. \eqno (45)
$$
 The transition to spherical coordinates is given by the relations
 $$
\delta_{ik}dx^idx^k=dr^2+r^2(d\theta^2+\sin^2\varphi d\varphi^2),\quad
\frac{x_ix_kdx^idx^k}{r^2}=(dr)^2.
 $$
The nonlinear correction $-4\phi^2$ in $g_{00}$ in (45) is of special interest
for us. The correct value necessary to explain the perihelion shift
is $+2\phi^2$ (if used in equation for geodesic, see. [19] and Sec. 6). The
 quantity $g_{00}$ is observable in principle by  redshift [16]
$\omega=\omega_0(g_{00})^{-\frac12}$. Why the nonlinear correction in (45)
 turns out
to be negative? It will appear later on that it is caused solely by the source
(15), see eqs. (95) and (117). The sources (8) and (15) have different signs,
but the corresponding fields have the same sign. The answer is simple. The
correction in (45) is only a small (and negative) part. The larger and
positive part goes for converting initially bare mass in Newtonian potential
into a dressed one, see eq. (113). Now the negative sign of correction is
clear: the mass of Newtonian center at infinitly large distance appears
as $M$, but at finite distance the test particle feels a greater mass and
greater attraction, because (15) is negative.

There are some technical problems on how to take into account  the
contributions from diagrams (2b), (2c) and Fig.1.
 To simplify the problem as much as possible
we may look only into corrections to Newtonian potential. As it will appear in
Sec.6, it seems impossible to get the right value on any assumption about the
contributions from these diagrams for the case of Thirring`s tensor.

We note here that Thirring  obtained from his tensor the necessary correction.
Yet his result is objectionable as he used illdefined gauge
 $$
\Box^2\Lambda=\frac{G^2M^2}{4r^2},
 $$
 see eq.  (83) in [2].  Namely the source
of $\Lambda$ fall of too slowly for large $r$ and the integral in (112), see
below, diverge for large $r'$.

In the next two Sections we shall see how energy-momentum "tensors" of general
relativity differ from Thirring`s tensor.

\section{ Landau-Lifshitz pseudotensor of energy-momentum}

This pseudotensor in the sense and approximation considered here is a tensor.
 In the lowest approximation with help of relation
 $$
\sqrt{-g}g^{ik}\approx(1+\frac h2)(\eta^{ik}-h^{ik})\approx%
\eta^{ik}-\bar h^{ik}. \eqno (46)
$$
we get from eq. (96.9) in [16]
\setcounter{equation}{46}
\begin{eqnarray}
t^{ik}=\frac{1}{16\pi G}[\bar h^{ik,l}\bar h_{lm}{}^{,m}-\bar%
h^{il}{}_{,l}\bar h^{km}{}_{,m}-\bar h^{km,p}\bar h_{mp}{}^{,i}-\bar%
h^{im,p}\bar h_{mp}{}^{,k}+ \bar h^{im,p}\bar h^k{}_{m,p}+ \nonumber \\
\frac12\bar h^{pq,i}\bar h_{pq}{}^{,k}-\frac14\bar h^{,i}\bar h^{,k}+\eta^{ik}%
(\frac12\bar h_{mn,p}\bar h^{pm,n}+\frac18\bar h^{,m}\bar h_{,m}-\frac14\bar h%
_{pq,m}\bar h^{pq,m})].
\end{eqnarray}
Comparison with canonical tensor (6) shows that it is connected with $t^{ik}$
by the relation
\setcounter{equation}{47}
\begin{eqnarray}
t^{ik}=\stackrel{f}{T}{}^{ik}+F^{ik},\quad \bar h_{ik}=-2f\varphi_{ik},
\nonumber\\
F^{ik}=\frac{1}{16\pi G}[\bar h^{ik,l}\bar h_{ln}{}^{,n}-\bar h^{il}{}_{,l}\bar
h^{kn}{}_{,n}- \bar h^{kn,p}\bar h_{np}{}^{,i}+\bar h^{in,p}\bar h^k{}_{n,p}].
\end{eqnarray}
 From (14) we see that now in place of $\stackrel{s}{T}{}^{ik}$  stands
 $F^{ik}$. But $\stackrel{s}{T}{}^{ik}$ was a conserved quantity, see.(29).
So  $F^{ik}$ should rather play the role of interaction energy-momentum tensor.
Indeed, taking into account that in considered approximation $\bar h_{ik}$
satisfies the linearized Einstein equation
 $$
 \bar h_{np,j}{}^j-\bar h_{jp,n}{}^j-\bar
h_{jn,p}{}^j+\eta_{np}\bar h_{qr}{}^{,qr}=-16\pi G{\cal T}_{np}, \eqno (49)
$$
we find
$$
F^{jk}{}_{,j}=\bar h^{kn,i}{\cal T}_{ni}=h^{kn,i}{\cal T}_{ni}-
\frac12h^{,i}{\cal T}_{i}{}^{k}. \eqno (50)
$$
Now we check that coservation laws [16]
$$
\frac{\partial}{\partial x^k}\left((-g)[\stackrel{p}{T}{}^{ik}+t^{ik}]\right)%
=0 \eqno (51)
$$
are fulfilled. From (48), (28) and (50) we have
$$
t^{ik}{}_{,i}=-\frac12h^{iq,k}{\cal T}_{iq}+h^{kn,i}{\cal T}_{ni}-\frac12h^{,i}%
{\cal T}_i{}^{k}. \eqno (52)
$$
For matter energy-momentum tensor from (17) we get
$$
(-g)\stackrel{p}{T}{}^{ik}=\sqrt{-g}{\cal T}^{ik}\approx(1+\frac h2){\cal T}%
^{ik},\quad -g\approx1+h.\eqno(53)
$$
From here
$$
\left((-g)\stackrel{p}{T}{}^{ik}\right)_{,i}={\cal T}{}^{ik}{}_{,i}+%
\frac12h_{,i}{\cal T}^{ik},\eqno(54)
$$
 the terms of order $h^2$ being dropped. Now it follows from (25), (52) and (54)
 that (51) is fulfilled. As seen from (53) here too there is a tensor, which
is nonzero only where particles are located. Surprisingly it coincides with
Thirring`s interaction tensor, see (37) and text below it.

Althouht  $-g\stackrel{p}T{}^{ik}+t^{ik}$ differs from Thirring`s
${\cal T}^{ik}+\stackrel{int}T{}^{ik}+\theta^{ik},$
 for Newtonian centers they
coincide. Now we turn to Newtonian approximation. According to Problem 1 in
 \S 106 in [16] the energy density of gravitational field in this approximation
is given by $\stackrel{f}{T}{}^{00}$ in (13), (10).  But there is also
energy density of interaction  $\mu\phi$, where $\mu$ is density of particles.
Using Poisson equation  $\nabla^2\phi=4\pi G\mu$ and ignoring the problems
connected with point-like nature of particles, we can write (utilizing
integration by parts)
 $$
\int\mu\phi dV=-\int\frac{1}{4\pi G}(\nabla\phi)^2dV,\eqno(55)
 $$
The density in the integrand on the r.h.s. contains now not only the energy
density of interaction, but also the proper energy density of particle`s
self-field. The density on the l.h.s. is nonzero only at particle locations,
the density on the r.h.s. is nonzero where the field is nonzero.
The integration by parts deprive us the possibility to retain the previous
physical meaning of integrand. If nevertheless we do this, then adding to (13)
the  energy density in the r.h.s. of (55) we get the effective gravitational
energy density in Newtonian approximation
 $$
 -\frac{1}{8\pi
G}(\nabla\phi)^2.\eqno(56)
 $$
 To bring this in agreement with  $t^{00}$ we ought, according to a foot-note
in  [16], take into account the contribution from
${(-g)\stackrel{p}{T}{}^{00}}$.  Let us do it. For $t^{00}$ we have
 $$
t^{00}=-\frac{7}{8\pi G}(\nabla\phi)^2,\eqno(57)
 $$
where $\phi$ is the potential of Newtonian centers. Now
 $$
(-g)\stackrel{p}{T}{}^{00}=\sqrt{-g}{\cal T}{}^{00}\approx{\cal T}^{00}%
(1+\frac h2) \approx{\cal T}^{00}-2\phi\mu,\eqno(58)
 $$
see (12). The sought for agreement will be reached only after we rewrite a la
Thirring [2] ${\cal T}{}^{00}$ in terms of observables. From (17)
 $$
 {\cal  T}^{00}=\sum_am_a\frac{dx^0}{ds}\delta(\vec x-\vec
x_a(t)).\eqno(59)
 $$
In the presence of gravitational field
 $$
ds^2=g_{00}dt^2(1-v^2).\eqno(60)
 $$
Here $v^2$ is physical velocity, see \S 88 in [16]. Hence
 $$
\frac{dx^0}{ds}=\frac{1}{\sqrt{g_{00}(1-v^2)}}\approx\frac{1}{\sqrt{1-v^2}}-
\frac{h_{00}}{2},\quad v^2\ll1. \eqno(61)
$$
Thus after going over to the observable velocity we obtain the term
 $$
-\frac{h_{00}}{2}\mu=-\phi\mu.\eqno(62)
 $$
 detached from $ {\cal T}{}^{00}$. Equation (12) was used to get  the r.h.s..
  Together with corresponding term in (58) this leads to
 $$
-3\mu\phi\Longrightarrow\frac{3}{4\pi G}(\nabla\phi)^2, \eqno(63)
 $$
where arrow corresponds to going over in (55) from integrand on the l.h.s.
to the integrand on r.h.s.. Now the sum of (57) and (63) gives the expected
 (56).

The consideration of Newtonian approximation makes the following point of view
very enticing: The energy density of an isolated point-like particle should be
positive; Hilbert gauge exclude the unnecessary spins and then positivity
seems quite natural, because the presence of virtual gravitons should not make
the energy density negative. The attraction is described by interaction energy
density and so the latter must be negative. Neither  Thirring tensor nor
Landau-Lifshitz one satisfies this requirement. The MTW tensor does.
Unfortunately I failed to fit this idea into existing approach to gravitation,
see Sec.6.

\section{ Papapetrou- Weinberg energy-momentum tensor.}

Einstein equation can be recast in such a way that gravitational energy
-momentum "tensor" can be easily identified in coordinate system that
goes over to Lorentzian at large distances from gravitating bodies [9]. In the
 lowest approximation this tensor have the form, see eq. (7.6.14) in
  [9])
  $$
 t_{\mu\kappa}=\frac{1}{8\pi%
 G}[-\frac12h_{\mu\kappa}R^{(1)}+\frac12\eta_{\mu\kappa}h^{\rho\sigma}R^{(1)}%
 _{\rho\sigma}+R^{(2)}_{\mu\kappa}-\frac12\eta_{\mu\kappa}R^{(2)}],\; R^{(1,2)}=
 R^{(1,2)}_{\mu}{}^{\mu}.
    \eqno(64)
 $$
In this Section we use Weinberg notation:
  $$
  \eta_{\mu\nu}=diag(-1,1,1,1),\;
 g_{\mu\nu}=\eta_{\mu\nu}+h_{\mu\nu},\; h_{\mu\nu}\equiv
 h^{W}_{\mu\nu}=-h^{Thir}=-h^{LL}_{\mu\nu}.\eqno(65)
  $$
 Greek indices run from 0 to 3, latin ones from 1 to 3. $R^{(1,2)}_{\mu\nu}$
 is Ricci tensor in the first and second approximation in powers of
  $h_{\mu\nu}$. The indices are raised and lowered with $\eta_{\mu\nu}.$ In
 terms of $\bar h_{\mu\nu}$ we get
  $$
  R^{(1)}_{\mu\nu}=\frac12(\bar
 h_{\mu\nu,\sigma}{}^{\sigma}-\bar h_{\sigma\mu,\nu}{}^{\sigma}-\bar
 h_{\sigma\nu,\mu}{}^{\sigma})-\frac14\eta_{\mu\nu}\bar
 h_{,\sigma}{}^{\sigma}\;,\; R^{(1)}=-\bar h^{\mu\sigma}%
 {}_{,\mu\sigma}-\frac12\bar h_{,\sigma}{}^{\sigma}\;, \eqno(66)
 $$
 \setcounter{equation}{66}
 \begin{eqnarray}
R^{(2)}_{\mu\kappa}=\frac12\bar h^{\lambda\nu}[\bar h_{\mu\nu,\kappa\lambda}+
\bar h_{\kappa\nu,\mu\lambda}-\bar h_{\lambda\nu,\kappa\mu}-\bar h_{\mu\kappa%
,\nu\lambda}]-\frac14(\bar h^{\lambda}{}_{\mu}\bar h_{,\kappa\lambda}+\bar
h_{\kappa}{}^{\nu}\bar h_{,\mu\nu}) \nonumber\\
+\frac14\bar h(\bar h_{,\mu\kappa}+\bar h_{\mu\kappa,\lambda}{}^{\lambda}-
\bar h_{\mu}{}^{\lambda}{}_{,\kappa\lambda}-\bar h^{\nu}{}_{\kappa,\nu\mu})-
\frac14(\bar h^{\nu}{}_{\mu,\nu}\bar h_{,\kappa}+\bar h^{\nu}_{\kappa,\nu}
\bar h_{,\mu})+\nonumber\\
\bar h^{,\sigma}(\frac12\bar%
h_{\mu\kappa,\sigma}-\frac14\bar h _{\kappa\sigma,\mu}-\frac14\bar%
h_{\mu\sigma,\kappa}) +\frac12\bar h^{\nu}{}_{\sigma,\nu}(\bar%
h^{\sigma}{}_{\mu,\kappa}+\bar h^{\sigma}{}_{\kappa,\mu}\nonumber\\
-\bar h_{\mu\kappa}{}^{,\sigma})+ \frac12\bar%
h_{\kappa\sigma,\lambda}(\bar h_{\mu}{}^{\lambda,\sigma}- \bar%
h_{\mu}{}^{\sigma,\lambda})-\frac14\bar h_{\sigma\lambda,\kappa} \bar%
h^{\sigma\lambda}{}_{,\mu}+\frac18\bar h_{,\mu}\bar h_{,\kappa}\nonumber\\ +
\eta_{\mu\kappa}[\frac14\bar h^{\lambda\nu}\bar h_{,\lambda\nu}-\frac18\bar h %
\bar h_{,\lambda}{}^{\lambda}+\frac14\bar h^{\nu}{}_{\sigma,\nu}\bar%
h^{,\sigma} -\frac18\bar h_{,\sigma}\bar h^{,\sigma}],
 \end{eqnarray}
\setcounter{equation}{67}
\begin{eqnarray}
R^{(2)}=\bar h^{\lambda\nu}(\bar h_{\kappa\nu}{}^{,\kappa}{}_{\lambda}-\frac12%
\bar h_{\nu\lambda,\sigma}{}^{\sigma})+\bar h^{\nu}{}_{\sigma,\nu}\bar %
h^{\sigma\kappa}{}_{,\kappa}-\frac12\bar h^{\nu}{}_{\sigma,\nu}\bar h^{,\sigma}%
\nonumber\\
+\frac12\bar h_{\kappa\sigma,\lambda}\bar h^{\kappa\lambda,\sigma}-\frac34%
\bar h_{\kappa\sigma,\lambda}\bar h^{\kappa\sigma,\lambda}-\frac12\bar h\bar h%
^{\kappa\lambda}{}_{,\kappa\lambda}+\frac18\bar h_{,\sigma}\bar h^{,\sigma}.
\end{eqnarray}
For Newtonian center from (9-12) we obtain
$$
\bar h_{\mu\nu}=-\bar h_{\mu\nu}^{T}=-4\phi\delta_{\mu0}\delta_{\nu0},\;
h_{\mu\nu}=-h_{\mu\nu}^{T}=-2\phi\delta_{\mu\nu},\; h=h^W=h^T=-4\phi=-\bar h.
\eqno(69)
$$
Nonzero components of $t_{\mu\kappa}$ are
$$
t_{00}=-\frac{3}{8\pi G}(\nabla\phi)^2=-\frac{3GM^2}{8\pi r^4},\; t_{ik}=
\frac{GM^2}{8\pi r^6}(7\delta_{ik}r^2-14x^ix^k).\eqno(70)
$$
In Hilbert gauge from equation
$$
\nabla^2\bar h_{\mu\nu}=-16\pi Gt_{\mu\nu}\eqno(71)
$$
 we find, cf. with (40), (42),
$$
\bar h_{00}=\frac{3G^2M^2}{r^2}=3\phi^2 ,\; \bar h_{ik}=-7G^2M^2\frac{x^ix^k}%
{r^4}.\eqno(72)
$$
 It is easy to check that (72) satisfies the Hilbert condition (2). In terms
of $h_{\mu\nu}$ we have
 $$
 h_{00}=-2\phi^2,\;
h_{ik}=G^2M^2(\frac{5\delta_{ik}}{r^2}-7\frac{x^ix^k}{r^4}).  \eqno(73)
 $$
In the expressions (71-73) $h_{\mu\nu}$ is nonlinear correction.

On the other hand in harmonic coordinates the Schwarzschild solution has the
form [9]
$$
-d\tau^2=-\frac{1+\phi}{1-\phi}dt^2+(1-\phi)^2(d\vec x)^2+\frac{1-\phi}{1+\phi}%
\phi^2\frac{x^ix^k}{r^2}dx^idx^k.\eqno(74)
$$
So in the considered approximation this gives
 $$
 g_{00}=-(1+2\phi+2\phi^2),\eqno(75)
 $$
 $$
g_{ik}=1-2\phi\delta_{ik}+\phi^2(\delta_{ik}+\frac{x_ix_k}{r^2}).\eqno(76)
$$
From (69) we have $h^{(1)}_{00}=-2\phi$, from (73) $h^{(2)}_{00}=-2\phi^2$,
and there is agreement with (75). As to the nonlinear correction for $g_{ik}$,
in (73) it differs from the one in (76) by a gauge. Really, subtracting from
 $h_{ik}$ in (73) the nonlinear part of (76), we find
 $$
G^2M^2(\frac{4\delta_{ik}}{r^2}-\frac{8x_ix_k}{r^4})=2G^2M^2(\Lambda_{i,k}+
\Lambda_{k,i}),\; \Lambda_i=\frac{x_i}{r^2},\eqno(77)
$$
i.e. a gauge.

Going back to $t_{00}$ in (70), we note that this density is negative
and does not coincide with any density of other tensors. At the same time
the equation of motion of particles is contained in the conservation laws of
total energy-momentum tensor. We shall check it in considered approximation.
For gravitational part the calculations give
 $$
 t^{\mu\kappa}{}_{,\kappa}=-h^{\nu}{}_{\sigma,\nu}{\cal
T}^{\mu\sigma}+\frac12%
 h_{,\sigma}{\cal
T}^{\mu\sigma}-\frac12h^{\rho\sigma,\mu}{\cal T}_{\rho\sigma}-
h^{\nu\lambda}{\cal T}^{\mu}{}_{\nu,\lambda}.\eqno(78)
$$
The energy-momentum tensor for particles, figuring in conservation laws,
has a rather complicated form by construction [9]
\setcounter{equation}{78}
 \begin{eqnarray}
\tau^{\mu\kappa}=\eta^{\mu\sigma}\eta^{\kappa\tau}g_{\sigma\alpha}g_{\tau\beta}%
\stackrel{p}{T}{}^{\alpha\beta}\approx\stackrel{p}{T}{}^{\mu\kappa}+h^{\mu}{}%
_{\alpha}{\cal T}^{\alpha\kappa}+h^{\kappa}{}_{\alpha}{\cal T}^{\alpha\mu}
\nonumber \\
\approx{\cal T}^{\mu\kappa}-\frac12h{\cal T}^{\mu\kappa}+h^{\mu}{}_{\alpha}%
{\cal T}^{\alpha\kappa}+h^{\kappa}{}_{\alpha}{\cal T}^{\alpha\mu}.
\end{eqnarray}
From here with the considered accuracy
$$
\tau^{\mu\kappa}{}_{,\kappa}=\stackrel{p}{T}{}^{\mu\kappa}{}_{,\kappa}+h^{\mu}%
{}_{\alpha,\kappa}{\cal T}^{\alpha\kappa}+h^{\kappa}{}_{\alpha,\kappa}%
{\cal T}^{\alpha\mu}+h^{\kappa}{}_{\alpha}{\cal T}^{\alpha\mu}{}_{,\kappa}.
\eqno(80)
$$
So
$$
(\tau^{\mu\kappa}+t^{\mu\kappa})_{,\kappa}=\stackrel{p}{T}{}^{\mu\kappa}%
{}_{,\kappa}-\frac12h^{\rho\sigma,\mu}{\cal T}_{\rho\sigma}+
 h^{\mu\alpha,\kappa}{\cal T}_{\alpha\kappa}+\frac12h_{,\sigma}{\cal T}%
 ^{\mu\sigma}.\eqno(81)
 $$
 Further from (53) we get
 $$
\stackrel{p}{T}{}^{\mu\kappa}{}_{,\kappa}\approx{\cal T}^{\mu\kappa}{}_{,\kappa}
- \frac12h_{,\kappa}{\cal T}^{\mu\kappa}.\eqno(82)
  $$
 Taking into account (25) we see that the r.h.s. of (81) is zero.

Now we turn to Newtonian approximation. Terms of interaction tensor are
  contained in both  $\tau^{\mu\nu}$ (three last terms
in the r.h.s. of (79)) and in
   $t_{\mu\kappa}$. From (64), (66-68), using (49), which preserve its form in
  the notation of this Section, we find the following terms of interaction
  tensor contained in $t_{\mu\kappa}$ :
  $$
  -\frac12h_{\mu\kappa}{\cal T}-\eta_{\mu\kappa}(\bar h^{\rho\sigma}{\cal T}%
  _{\rho\sigma}-\frac14\bar h{\cal T})-\frac12\bar h{\cal T}_{\mu\kappa}.
  $$
 From here and the first equation in (70) we get in Newtonian approximation
    $$
   t_{00}=-\frac{3}{8\pi G}(\nabla\phi)^2-6\phi{\cal T}_{00}.\eqno(83)
   $$
  Here $\phi$ is the same as in (10). From (79) and (53) we get in this
  approximation
$$
   \tau^{00}=(-g)^{-\frac12}{\cal T}^{00} +2h^{0}{}_{0}{\cal T}^{00}\approx%
{\cal T}^{00}(1-\frac12h)-2h_{00}{\cal T}^{00}={\cal T}^{00}+6\phi{\cal T}%
  ^{00}.\eqno(84)
   $$
  Thus in Newtonian approximation the interaction terms in the sum of (83) and
 (84) are cancelled out. The agreement with Newtonian approximation (56) is
achieved in the same way as for Landau-Lifshitz tensor: ${\cal T}^{00}$ on the
r.h.s. of (84) detaches term (62), which is equivalent (in accordance with
(63)) to $\frac{1}{4\pi G}(\nabla\phi)^2$. Together with the first term on the
 r.h.s.  of (83) this gives (56).

Weinberg shows in detail that his energy-momentun tensor has all requiered
characteristics. But this tensor does not help us to find energy-momentum
tensor of two gravitons as represented by straight line on diagrams of Fig.2.
By construction Weinberg`s tensor gives the gravitational field only via
diagram (2a). The field-theretical description tell us that test particle
is not quite passive. It does not simply follows the command "move along
geodesic" but itself takes part in the creation of field in which it moves,
see Fig.1 and Fig.(2b, 2c). From this viewpoint one can expect that e.g.
photon and graviton scatter differently on Newtonian center as only in the
latter case all three diagrams of Fig.2 contribute.

\section{ Misner-Thorne-Wheeler energy-momentum tensor}

In this Section it is handy for us to use again Thirring`s notation. Up to
divergence terms the Lagrangian (6) may be rewritten as [6]
 $$
 {\cal
L}=\frac12\varphi_{\mu\nu,\lambda}\varphi^{\mu\nu,\lambda}-\frac14\varphi%
_{,\lambda}\varphi^{,\lambda}-\varphi_{\mu\nu}{}^{,\mu}\varphi^{\lambda\nu}%
{}_{,\lambda}.\eqno(85)
$$
The corresponding canonical energy-momentun tensor
 $$
\stackrel{f}{T}{}^{jk}=\frac{\partial{\cal L}}{\partial\varphi^{\mu\nu}
{}_{,j}}\varphi^{\mu\nu,k}-\eta^{jk}{\cal L},\eqno(86)
 $$
acquires the form
 $$
 \stackrel{f}{T}{}^{jk}=\bar T^{jk}-\frac12\eta^{jk}\bar
T\;,\bar T^{jk}=
\varphi^{\mu\nu,k}\varphi_{\mu\nu}{}^{,j}-\frac12\varphi^{,k}\varphi^{,j}-
2\varphi^{j\nu,k}\varphi_{\nu\sigma}{}^{,\sigma}\;, \bar T =-T=2{\cal L}.
\eqno(87)
$$
Spin part is given by (31-32) with substitution $L\to{\cal L}$. We dwell on
differences from Thirring`s tensor. In symmetric in $jk$ tensor
\setcounter{equation}{87}
 \begin{eqnarray}
 F^{jik}+F^{kij}=(\varphi^{\alpha
i,j}-\varphi^{\alpha\sigma}{}_{,\sigma}\eta%
^{ji})\varphi^{k}{}_{\alpha}+(\varphi^{\alpha i,k}-\varphi^{\alpha\sigma}{}%
_{,\sigma}\eta^{ki})\varphi^{j}{}_{\alpha}\nonumber \\
-2\varphi^{i\sigma}{}_{,\sigma}\varphi^{kj}+(2\varphi^{\alpha\sigma}{}_{,\sigma}%
\eta^{jk}-\varphi^{\alpha k,j}-\varphi^{\alpha j,k})\varphi^i{}_{\alpha}+
\varphi%
^{k\sigma}{}_{,\sigma}\varphi^{ij}+\varphi^{j\sigma}{}_{,\sigma}\varphi^{ik}
\end{eqnarray}
there is no derivatives over $x^i$. This means that in divergence
 $F^{jik}{}_{,i}+%
 F^{kij}{}_{,i}$ there are no terms of interaction tensor. In antisymmetric
in  $jk$ tensor
 $$
 F^{ikj}=(\varphi^{\alpha
k,i}-\varphi^{\alpha\sigma}{}_{,\sigma}\eta^{ik})%
\varphi^{j}{}_{\alpha}-\varphi^{k\sigma}{}_{,\sigma}\varphi^{ij}+(\varphi%
^{\alpha\sigma}{}_{,\sigma}\eta^{ij}-\varphi^{\alpha j,i})\varphi^k{}{}_%
{\alpha}+\varphi^{j\sigma}{}_{,\sigma}\varphi^{ik}\eqno(89)
$$
such terms are present. Hence, using linearized Einstein equation (27) in the
expression for $F^{ikj}{}_{,i}$, we get
 $$
 -F^{ikj}{}_{,i}=f(\bar{\cal
T}^{j\alpha}\varphi^{k}{}_{\alpha}-\bar{\cal T}%
^{k\alpha}\varphi^{j}{}_{\alpha})+\varphi^{\alpha\sigma}{}_{,\sigma}(\varphi%
^{j}{}_{\alpha}{}^{,k}-\varphi^{k}{}_{\alpha}{}^{,j}).\eqno(90)
$$
Terms with $f$ together with (30) give symmetric interaction tensor in
accordance with (34). Other two  terms on the r.h.s. of (90)
supplement $\stackrel{f}{T}{}^{jk}$ to symmetric one, see (87). So we get
 \setcounter{equation}{90}
\begin{eqnarray}
\theta^{jk}=\stackrel{f}{T}{}^{jk}+\stackrel{s}{T}{}^{jk}=\varphi^{\mu\nu,k}%
\varphi_{\mu\nu}{}^{,j}-\frac12\varphi^{,k}\varphi^{,j}-\varphi^{j\sigma}{}%
_{,\sigma i}\varphi^{ik}-\varphi^{k\sigma}{}_{,\sigma i}\varphi^{ij}\nonumber\\
-\varphi^{\alpha i,j}\varphi^{k}{}_{\alpha,i}-\varphi^{\alpha i,k}\varphi^{j}%
{}_{\alpha,i}+(\varphi^{\alpha j,k}{}_{i}+\varphi^{\alpha k,j}{}_{i})\varphi%
^{i}{}_{\alpha}+2\varphi^{i\sigma}{}_{,i\sigma}\varphi^{jk}+\nonumber\\
2\varphi^{i\sigma}{}_{,\sigma}\varphi^{kj}{}_{,i}-2\varphi^{j\sigma}{}_{,\sigma}%
\varphi^{ki}{}_{,i}+(\varphi^{\alpha k,j}+\varphi^{\alpha j,k})\varphi_%
{\alpha\sigma}{}^{,\sigma}\nonumber\\
-2\eta^{jk}(\varphi^{\alpha\sigma}{}_{,\sigma}\varphi^i{}_{\alpha})_{,i}-
\eta^{jk}{\cal L}+f(\bar{\cal T}^{j\alpha}\varphi^k{}_{\alpha}-\bar{\cal T}
^{k\alpha}\varphi^j{}_{\alpha}).
\end{eqnarray}
From (91) and (30), using the relation between $\varphi_{\mu\nu}$ and
 $\bar h_{\mu\nu}$ in (9), and taking into account (34), we find
\setcounter{equation}{91}
\begin{eqnarray}
\theta^{jk}+\stackrel{int}{T}{}^{jk}=\frac{1}{32\pi G}[\bar h^{\mu\nu,k}%
\bar h_{\mu\nu}{}^{,j}-\frac12\bar h^{,k}\bar h^{,j}- (\bar h^{j\sigma}{}%
_{,\sigma i}\bar h^{ik}+\bar h^{k\sigma}{}_{,\sigma i}\bar h^{ij})\nonumber\\
-(\bar h^{\alpha i,j}\bar h^k{}_{\alpha,i}+\bar h^{\alpha i,k}\bar
h^j{}_{\alpha,i})+(\bar h^{\alpha j,k}{}_{i}+\bar h^{\alpha k,j}{}_{i})\bar h%
^i{}_{\alpha}+2\bar h^{i\sigma}{}_{,\sigma i}\bar h^{jk}+2\bar h^{i\sigma}{}%
_{,\sigma}\bar h^{kj}{}_{,i} \nonumber\\
-2\bar h^{j\sigma}{}_{,\sigma}\bar h^{ki}{}_{,i}+(\bar h^{\alpha k,j}+
\bar h^{\alpha j,k})\bar h_{\alpha\sigma}{}^{,\sigma}-2\eta^{jk}(\bar
h^{\alpha\sigma}{}_{,\sigma}\bar h^i{}_{\alpha})_{,i}-\eta^{jk}{\cal
L}]\nonumber\\
+\frac12({\cal T}^{k\alpha} h_{\alpha}{}^j+{\cal
T}^{j\alpha} h_{\alpha} {}^k),
 \end{eqnarray}
 $$
{\cal L}=\frac1{32\pi G}[\frac12\bar h_{\mu\nu,\lambda}\bar h^{\mu\nu,\lambda}-
\frac14\bar h_{,\lambda}\bar h^{,\lambda}-\bar h_{\mu\nu}{}^{,\mu}\bar
h^{\lambda\nu}{}_{,\lambda}].\eqno(93)
$$
It is easy to verify that total energy-momentum tensor consisting of (17) and
(92) is conserved.

For Newtonian centers from (92) and (9), (12) we have
 $$
 \theta^{00}+\stackrel{int}{T}{}^{00}=\frac1{8\pi
G}(\nabla\phi)^2+2\mu\phi, \;\mu={\cal T}^{00}.\eqno(94)
 $$
For this system the MTW Lagrangian (93) coincide with Thirring`s one. The same
is true for canonical energy-momentum tensors, see (6) and (86-87), but spin
parts are different. It follows from (91) and (87) that for Newtonian
centers MTW spin part contributes only to interaction tensor, while
Thirring`s spin part contributes also to pure field part, see (15).

We note now that in Hilbert gauge for static case (for Newtonian
centers) the linearized Einstein equation can be written as
 $$
\nabla^2h^W_{\mu\nu}=-\nabla^2h^T_{\mu\nu}=-16\pi G\bar T_{\mu\nu},\eqno(95)
 $$
see (116) below. As seen from (87) for this system  $\bar T_{00}=0$, i.e.
there is no contribution to $h_{00}$  from diagrams of Fig.2.

Comparing MTW and Thirring tensors in Newtonian approximation we see that
addition of divergence terms to Lagrangian leads to the change in subdivision
of energy density between purely field part and interaction part.

\section{ On Schwinger`s explanation of perihelion shift }

Schwiger gave an elementary explanation of perihelion precession [10]. His
 method is of interest in many respects. The offered algorithm for calculating
 gravitational field is more in line with  field-theoretical method than finding
 field via differential equation in geometrical approach. In linear
approximation both methods naturally agree. In higher approximation
   Schwinger`s gravitational field, defined as a coefficient at
   $\delta T^{\mu\nu}$ in variation of
amplitude
  $$ \delta W(T)=\int d^4x\delta T^{\mu\nu}h^{Sch}_{\mu\nu},\quad
\delta T^{\mu\nu}{}_{,\nu}=0, \quad h_{\mu\nu}=2h^{Sch}_{\mu\nu},\eqno(96)
  $$
 seems does not coincide with definition of $h_{\mu\nu}$ in (1), because
 the nonlinear correction to  $\phi$ is $\frac12\phi^2$, but not  $\phi^2$,
 see the text below eq.(108).  In the lowest approximation
  \setcounter{equation}{96}
\begin{eqnarray} W(T)=4\pi G\int
d^4xd^4x'T^{\mu\nu}(x)D_{\mu\nu\rho\sigma}(x-x')T^{\rho\sigma}
 (x'),\;D_{\mu\nu\rho\sigma}=P_{\mu\nu\rho\sigma}D_{+}(x-x'),
\end{eqnarray}
$$
D_+(x)= \frac i{(2\pi)^3}\int\frac{d^3p}{2p^0}
e^{i(\vec p\cdot\vec x-p^0\vert t\vert)},
$$
$$
P_{\mu\nu\rho\sigma}=\frac12(\eta_{\mu\rho}\eta_{\nu\sigma}+\eta_{\mu\sigma}%
\eta_{\nu\rho}-\eta_{\mu\nu}\eta_{\rho\sigma}).\eqno(98)
$$

The amplitude $W$ is related to the energy of system $E$ by the expression
 $$
 W(T)=-\int dtE(t).\eqno(99)
 $$
From the total energy $E$ Schwinger singles out the interaction energy
 $E^{int}$. Nonlinear correction to the interaction energy of slowly moving
planet with the Sun is given then by
 $$
\stackrel{(2)}{E}{}^{int}=-GM\int\frac{d^3x}{\vert\vec x\vert}t^{00 int},
\eqno(100)
 $$
where $t^{00 int}$ is taken in Newtonian approximation
 $$
 t^{00 int}=-\frac{GMm}{4\pi}\nabla\frac1{\vert\vec
x\vert}\cdot\nabla\frac1 {\vert \vec x-\vec R\vert}.\eqno(101)
 $$
So the result is
 $$
\stackrel{(2)}{E}{}^{int}=\frac m2\phi^2(x),\quad \phi(x)=-\frac{GM}{R}.
\eqno(102)
 $$

Eq.(100) follows from eq.(4.34) in Ch.2 in [10] if one assums that
energy-momentum tensor of gravitational field is given by  1/8 of the
expression (15), in which $\phi$ is that of (10), i.e. the potential of
 Sun-planet system. The Sun`s matter energy-momentum tensor is represented by
(8). It is clear that (100) is given by contribution from diagrams  (2b) and
(2c).

As shown in Sec.3 the energy density of gravitational field may be cosidered
as positive. In that case the expression for $\stackrel{(2)}{E}{}^{int}$ in
(102) changes sign, but then there is also a contribution from interaction
energy density
 $$
 \mu\phi=-\frac{GMm}{R}[\delta(\vec x)+\delta(\vec
x-\vec R)].\eqno(103)
 $$
Using this instead of $t^{00int}$ in (100) and dropping the contribution from
 $\delta(\vec x)$ (self-interaction is included in renormalization of Sun`s
mass) we get
 $$
\stackrel{(2)}{E}{}^{int,loc}=\frac{G^2M^2m}{R^2}=m\phi^2.\eqno(104)
 $$
This together with (102), taken with reversed sign, restores Schwinger`s
result. We see that in this way the explanation of perihelion shift is
achieved without the concept of negative energy density of gravitational field.

So the nonlinear correction to the potential is taken into account by
substitution
 $$
 \phi\to\phi^{eff}=\phi(1+\frac12\phi).\eqno(105)
 $$
Therefore the Newtonian attraction is decreased by this correction.
 Qualitatively this result is quite understandable: the correction term
 describes the interaction of the Sun with negative interaction energy and this
 corresponds  to repulsion. The effective potential (105) produces
 the acceleration
  $$
  -\frac{%
 d}{dr}\phi^{eff}=-\phi'(1+\phi)=-\frac{GM}{r^2}(1-\frac{GM}r).\eqno(106)
   $$

 In general relativity the expression for acceleration of particle at rest
 depends upon the chosen coordinates. In Schwarzschild coordinates we have
  $$
 \sqrt{F_iF^i}=\frac{GM}{R^2\sqrt{1-\frac{r_g}R}},\quad r_g=2GM,\eqno(107)
  $$
 see e.g. eqs. (A 61), (A 63) in Appendix in [13]. In harmonic coordinates
  $r=R-\frac{r_g}2$ the expression takes the form
   $$
  \sqrt{F^iF_i}=\frac{GM}{r^2(1+\frac{r_g}{2r})\sqrt{1-(\frac{r_g}{2r})^2}}=
  \frac{GM}{r^2}(1-\frac{GM}r+\cdots),\eqno(108)
  $$
and does not contradict (106). We note also that in this coordinates  $g_{00}$
is given by (75), i.e. the Newtonian potential $\phi$ is substituted
by $\phi(1+\phi)$. The difference with (105) is caused by the fact that
Schwinger  obtained the correction to $\phi$ for  use in Lagrangian method,
cf. \S 106 in [16] and [21], not for correcting $g_{00}$.  The relation is simple. Let us denote by $\frac12\alpha$ and $\beta$
the coefficients at $\phi^2$ in corrections to $\phi$ in $g_{00}$ and in
Schwinger method.  Then  from eq. below (106.15) in [16] we have for particle
at rest $$ L \propto \frac{ds}{dt}=\sqrt{1+h_{00}}\approx 1+\phi+
\left(\frac{\alpha}{2}-\frac12\right)
\phi^2,
$$
where
$$
h_{00}=1+2\phi+\alpha\phi^2.
$$
So $\beta=\frac{\alpha}{2}-\frac12.$

Now we shall see what is in store for us if we use the diagram method. First
we dwell on properties of graviton propagator (98). The polarization factor
 $P_{\mu\nu\rho\sigma}$  satisfies the relations
 $$
t^{\mu\nu}P_{\mu\nu\rho\sigma}=t_{\rho\sigma}-\frac12\eta_{\rho\sigma}t\equiv
\bar t_{\rho\sigma} ,\quad P_{\mu\nu\rho\sigma}T^{\rho\sigma}=\bar T_{\mu\nu},
$$
$$
t^{\mu\nu}P_{\mu\nu\rho\sigma}T^{\rho\sigma}=t^{\mu\nu}T_{\mu\nu}-\frac12tT=
t^{\mu\nu}\bar T_{\mu\nu}=\bar t^{\mu\nu}T_{\mu\nu}.\eqno(109)
$$
The scalar factor  $D_+(x)$ has the representation
$$
D_+(x)=\frac{1}{4\pi}\delta_+(x^2)=\frac{i}{(2\pi)^2}\frac{1}{x^2+i\epsilon},
\eqno(110)
$$
and possesses the property
$$
\int d\tau D_+(\vec x-\vec x',\tau)=\frac i{(2\pi)^2}\int^{\infty}_{-\infty}
\frac{d\tau}{(\vec x-\vec x')^2-\tau^2+i\epsilon}=\frac1{4\pi\vert \vec x-\vec
x' \vert }.\eqno(111)
 $$

For spherically symmetric body we have to deal  with integrals of the kind
 $$
  \int d^4x'D_+(x-x')\rho(r')=\frac1{4\pi}\int\frac{d^3x'}{\sqrt{\vec
x'^2+\vec x^2-2\vec x\cdot\vec
  x'}}\rho(r')=\frac1r\int^r_0dr'r'^2\rho(r')+\int^{\infty}
  _rdr'r'\rho(r'),\eqno(112)
  $$
By the way it is seen from here that the derivative of Newtonian
potential over $r$ is determined only by the mass inside sphere of
  radius $r$. Assuming in (112) that  $\rho=\frac c{r^4}$, we get
 $$
  \frac1r\int^r_{\delta}dr'r'^2\rho(r')
+\int^{\infty}_rdr'r'\rho(r')=c\left(\frac1{r\delta} -\frac1{2r^2}\right).
\eqno(113)
 $$
Hence the divergent part at small $r'$ appears only in the term,
which is absorbed by mass renormalization.

So the source (13) generate the field
$$
\bar h_{00}=16\pi G\int d^4x'D_+(x-x')T_{00}(x')\Longrightarrow-\phi^2.
\eqno(114)
$$
The arrow  shows that the divergent part  is included in mass
renormalization. The r.h.s. of (114) can be obtained also from the
solution of wave equation derived from (114) by action of the operator
 $\partial^2=\nabla^2- \frac{\partial^2}{\partial t^2} $ and taking into
account that
 $$
 -\partial^2D_+(x-x')=\delta(x-x')\;,\quad
\nabla^2\frac1{r^2}=\frac2{r^4}.  \eqno(115)
 $$
We note also that
 $$
 h_{\mu\nu}=2h^{Sch}_{\mu\nu}=16\pi G\int
d^4x'D_+(x-x')\bar T_{\mu\nu}(x').  \eqno(116)
 $$

First we try MTW energy-momentum tensor as a source of gravitational field.
 The canonical part of MTW tensor in Hilbert gauge has the
form
 \setcounter{equation}{116}
 \begin{eqnarray}
\stackrel{f}{T}{}^{\gamma\delta}=\frac1{32\pi G}\{\bar h^{\mu\nu,\delta}
\bar h_{\mu\nu}{}^{,\gamma}-\frac12\bar h^{,\delta}\bar h^{,\gamma}\nonumber\\
-\eta
^{\gamma\delta}[\frac12\bar h_{\mu\nu,\lambda}\bar h^{\mu\nu,\lambda}-
\frac14\bar h^{,\lambda}\bar h_{,\lambda}]\}=\bar T^{\gamma\delta}-\frac12\eta
^{\gamma\delta}\bar T.
\end{eqnarray}
For Newtonian centers this expression coincides with Thirring`s
one, see eq.(6). As seen from (117) in this static case  $\bar T^{00}=0$. So
there is no contribution to $h_{00}$ from this source. As
noted in Sec.5 the nonlocal part of $\stackrel{s}{T}{}^{\mu\nu}$ is zero for
Newtonian centers. So assuming, as in the case of Weinberg tensor,
that the contribution from diagrams of Fig.1 and (2b), (2c) should
not be taken into account, we cannot obtain $g_{00}$ necessary for
explaining the perihelion shift; even a combination of MTW and Thirring tensors
will not help in this case.

Thus we have to take into account all diagrams. As we know from general
relativity, where the vertices are known, the extracting post-Newtonian
corrections from scattering amplitudes is a cumbersome and labour consuming job
[21, 22].  For this reason  we assume here the Schwinger method.

Returning to Thirring tensor we find that the total correction to $\phi$ from
all diagrams of Fig.1-2 is zero. Indeed starting from the
    energy density (56), Schwinger obtained the correction to $\phi$ as
 $\frac12\phi^2$. Therefore the corresponding contribution from spin
part (15)
    is $4\phi^2$. This correction comes from diagrams (2b) and (2c).
 Similarly to (103), (104) we find that the correction
to $\phi$ from diagram of Fig.1 is $-2\phi^2.$ Finally the diagram (2a),
treated according to Schwinger method, contributes $-2\phi^2.$ So we get zero
instead of $\frac12\phi^2$, necessary for use in Lagrangian method, see \S 106
in [16].

 Now it easy to see that a linear combination of MTW and Thirring tensors
with weights $\frac14$ and $\frac34$ gives the necessary correction
 $\frac12\phi^2.$ As the MTW tensor alone is unable to explain the perihelion
shift, we have to give up the tempting disire to consider the gravitational
energy density of an isolated point-like center as positive.

Besides the nonlinear correction (102) there is  a nonrelativistic linear
    correction to the Sun-planet interaction energy. All nonrelativistic
    corrections have the form [10]
     $$
     \frac{3\vec
    p^2}{2m^2}V+\frac1{2m}V^2\;,\quad V=m\phi.\eqno(118)
     $$
    Relativistic corrections may be treated either according to Schwinger or
    by using (118) together with Newtonian potential energy  $V$ in
    relativistic equation of motion in flat space,  see i.g. Ch.5, \S
    1 in [20]. Both methods give the correct result. An elaborate analysis
     of different aspects of planet motion in this approximation is given in
    [19]. Finally we remind that three other famous observational effects
    of general relativity are explained already by linear approximation
     [2,10,19].
    \section{ Concluding remarks}

   Though general covariance was not assumed, the gauge invarians is
    of course retained [2,10]. For this reason the weak gravitatinal waves
    in flat space are described as in general relativity. All considered
    tensors give  the same energy-momentum tensor for the plane gravitational
    wave. There are no a priori reasons to believe that field-theoretical
    approach will give the same result as general relativity. Similar assertion
    was made in [21]. Our investigation shows that there is still
    much to be done to synthesize the geometrical and field-theoretical aspects
of gravitations.

      I wish to thank V.I.Ritus, S.L.Lebedev, I.V.Tyutin, and M.A.Vasiliev
    for useful discussions. I am particularly indebted to V.Ya. Fainberg
    for remarks contributing to improvement of the manuscript.
   This work was supported by the Russian Fund for Fundamental Research
    (Grants No 96-02-17314  and 96-15-96463).

\begin{center}  Figure captions \end{center}

Fig.1: The second rank tensor formed from matter energy-momentum tensor and
graviton is a source  for another graviton.

Fig.2: 3-graviton vertex. Short straight line serves only to distinguish the
roles of participating gravitons: energy-momentum tensor is formed from
two gravitons joining the straight line at its ends, this energy-momentum
  tensor
serves as a source for graviton emerging from the middle of the straight line.
Crosses represent external field sources.  \section{References} 1.
W.E.Thirring, Fortschr. Physik. {\bf7}, 79 (1959). \\ 2. W.E.Thirring, Ann.
Phys. (N.Y.) {\bf16}, 96 (1961).  \\ 3. V.I.Ogievetsky and I.V.Polubarinov,
Ann. Phys. (N.Y.) {\bf35}, 167 (1965).\\ 4. S.N.Gupta, Rev. Mod. Phys. {\bf29},
334 (1957).      \\ 5. S.S.Gerstein, A.A.Logunov, and M.A.Mestvirishvili, Dokl.
Akad. Nauk, Russia, Vol.  {\bf360}, p.  332 (1998) (in Russian).\\ 6.
C.W.Misner, K.S.Thorne, J.A.Wheeler, {\sl Gravitation.} San Fracisco(1973).\\ 7.
  E.Alvarez, Rev. Mod.  Phys. {\bf61}, 561 (1989).\\ 8.  Yu.V.Graz, V.Ch.
  Zhukovski, and D.V.Galtsov, {\sl Classical fields}, Moscow (1991), Moscow
university publishing hous (in Russian). \\ 9.  S.Weinberg, {\sl Gravitation
  and Cosmology}, New York (1972).\\ 10.  J.Schwinger, {\sl Particles, Sources,
 and Fields.} V.1 Addison-Wesley (1970).\\ 11.  A.Pais, {\sl The Science and
  the Life of Albert Einstein}, Oxford (1982).  \\ 12.  A.Einstein, Ann. Math.
  {\bf40}, 922 (1939).\\ 13.I.D.Novikov, V.P.Frolov, {\sl Physics of black
  holes}, Moscow (1986) (in Russian).\\ 14.  Venzo de Sabbata and Maurizio
  Gasperini, {\sl Introduction to Gravitation.} World Scientific (1985).\\ 15.
  H.Dehnen, Zeitschr. f\"ur Phys.  {\bf179}, 1 Heft, 76 (1964).  \\
  16.L.D.Landau and E.M.Lifshitz, {\sl The
 classical theory of fields}, Moscow, (1973) (in Russian).  \\
  17. H.Weyl {\sl
 Raum, Zeit, Materie.} Springer Verlag, Berlin (1923).\\
  18.  H.Umezawa, {\sl
Quantum Field Theory}, Amsterdam (1956).\\
  19.H.Dehnen, H.H\"onl, and
K.Westpfahl, Ann.  der Phys. {\bf6}, 7 Folge, Band 6, Heft 7-8, S.670 (1960).
\\
 20.  A.Sommerfeld {\sl Atombau und Spectrallinien}, Band 1, Braunschweig
 (1951).\\
  21.Y.Iwasaki, Prog. Theor. Phys.  {\bf46}, 1587 (1971). \\
 22. H.W.Hamber, S.Liu. Phys. Lett. {\bf B 357}, 51 (1995).

 E-mail: nikishov@lpi.ac.ru

\newpage

\epsfxsize=\hsize
\hbox to\hsize{\hss\epsfbox{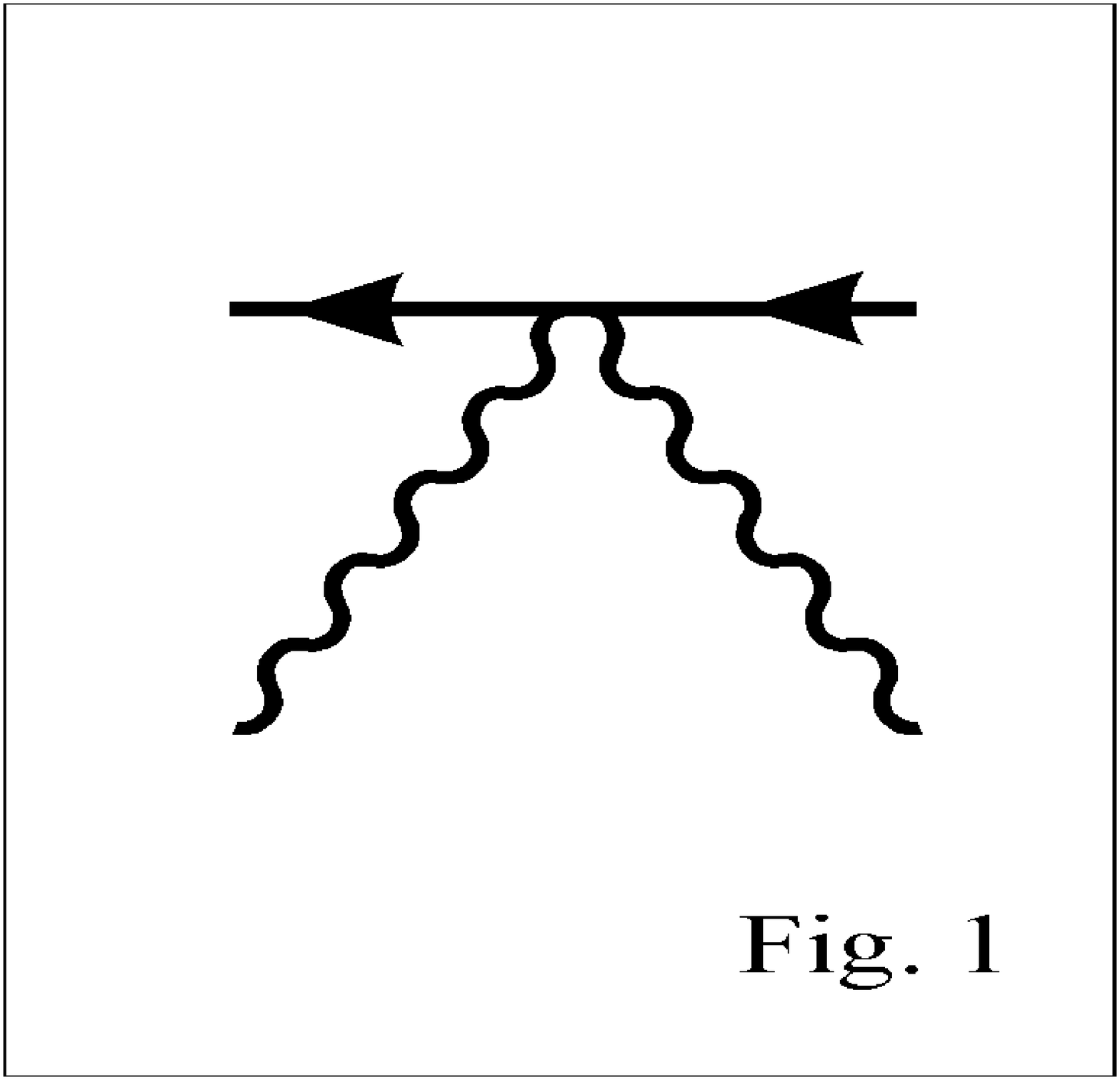}\hss}

\vskip 1in

\hbox to\hsize{\hss\epsfxsize=\hsize \epsfbox{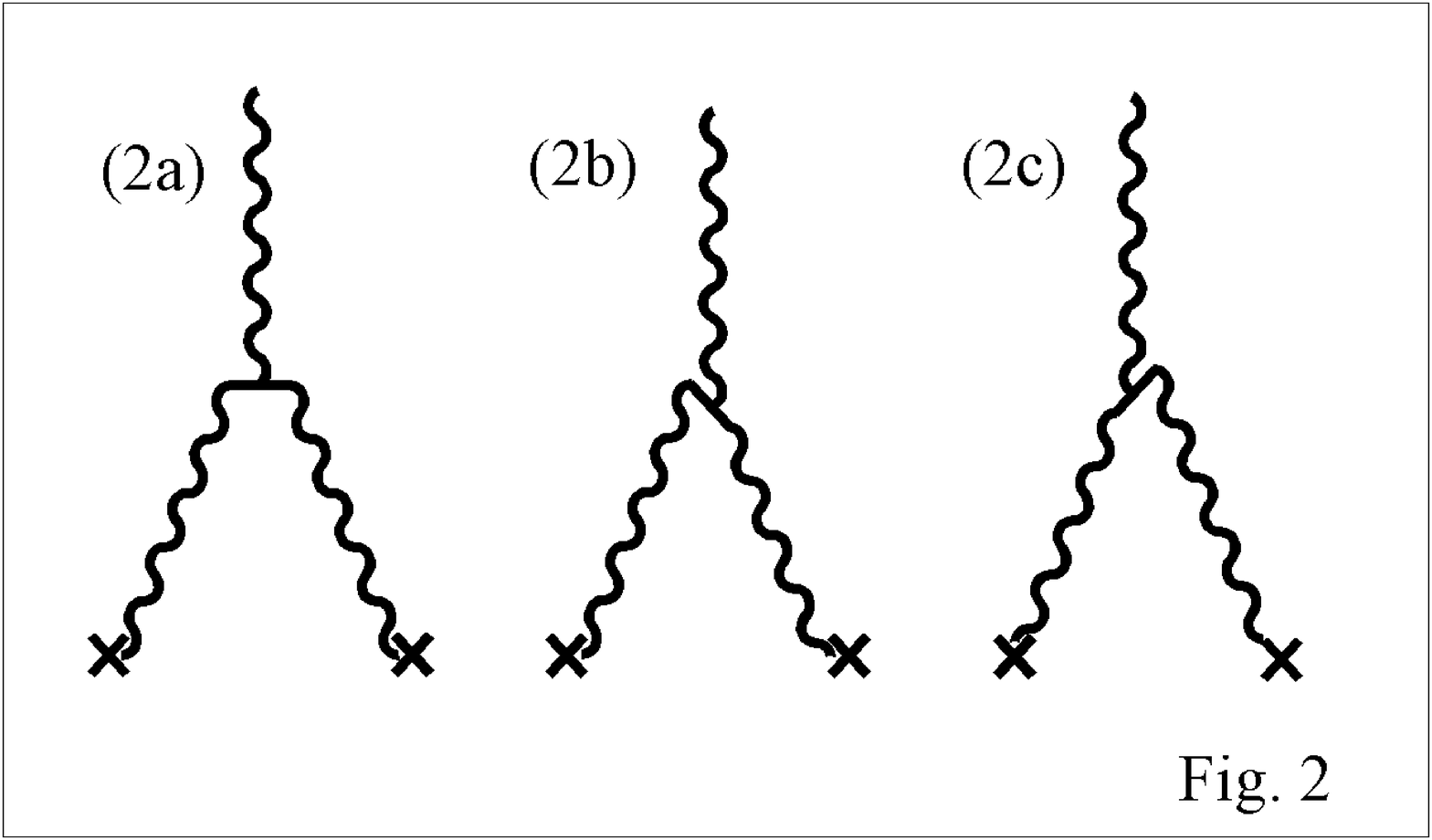}\hss}

\end{document}